\documentclass[onecolumn,showpacs,preprintnumbers,superscriptaddress,
amsmath,amssymb,floatfix,aps]{revtex4}

\usepackage[dvips]{graphicx}
\usepackage{amsmath, latexsym,  amssymb, array, graphics, eucal,
amsfonts, amsthm, amsmath,  amsfonts, array, bm}
\newcommand{\const}{\mathop{\rm const\, }}
\begin{document}
\newcommand{\mc}[1]{\mathcal{#1}}
\newcommand{\E}{\mc{E}}
\topmargin=5mm
\large

\title{Longitudinal %electric conductivity and dielectric
permeability of %quantum non--degenerate
collisional plasmas
under arbitrary degree of degeneration of electron gas}

\author{\bf A. V. Latyshev}
\affiliation{Department of Mathematical Analysis and Geometry,
Moscow State Regional
University,  105005,\\ Moscow, Radio st., 10--A}
\email{avlatyshev@mail.ru}

\author{\bf A. A. Yushkanov}
\affiliation{Department of Theoretical Physics,
Moscow State Regional
University,  105005,\\ Moscow, Radio st., 10--A}
\email{yushkanov@inbox.ru}

\date{\today}

\begin{abstract}\large
Electric conductivity and dielectric permeability of the non--degenerate
electronic gas for the collisional plasmas
under arbitrary degree of degeneration of electron gas is found.
The kinetic equation of  Wigner --- Vlasov --- Boltzmann
with collision integral  in relaxation form BGK (Bhatnagar, Gross
and Krook) in coordinate space is used.
%We will notice that
Dielectric permeability with using of the
relaxation equation in the momentum  space  has been received by Mermin.
Comparison with Mermin's formula has been realized.
It is shown, that in the limit when Planck's constant tends to
zero expression for dielectric permeability passes
in the classical.

\end{abstract}
\date{\today}
\maketitle

\section{Introduction}

In the present work formulas  for electric
conductivity and for dielectric permeability of quantum
electronic plasma under arbitrary degree of degeneration
of electron gas are deduced.

Dielectric permeability is one of the major
plasma characteristics. This quantity is necessary for
description of process of propagation and attenuation of the plasma
oscillations, skin effect, the mechanism of
electromagnetic waves penetration
in plasma \cite{Shukla1} -- \cite{Silin-Ruh}, and for
analysis of other problems in plasma physics.

Dielectric permeability in the collisionless quantum
gaseous plasma was studied by many authors (see, for example,
\cite {Klim}-\cite {Arnold1}). In work \cite {Manf}, where
the one--dimensional case of the quantum plasma is investigated,
importance of derivation of dielectric permeability with
use of the quantum kinetic equation with collision integral
 in the form of BGK -- model (Bhatnagar, Gross, Krook)
\cite {BGK} was marked.
The present work is devoted to performance of this  problem.

In the present work for a derivation of dielectric permeability
 the quantum kinetic Wigner --- Vlasov --- Boltzmann equation
(WVB--equation) with  collision integral
in the form of $ \tau $--models is applied.
Such collision integral is named BGK--collision integral.

The WVB--equation is written for   Wigner function,
which is analo\-gue of a distribution function of
electrons for quantum plasma
(see \cite{Wigner}, \cite{Hillery} and \cite{Kozlov}).

The most widespread method of investigation of quantum plasmas is the
method of Hartree --- Fock  or a method equivalent to it,
namely, the method of Random Phase Approximation
\cite {Lif}, \cite {Plaz}.
In work \cite {Lind} this method  has been applied to receive expression for
dielectric permeability of quantum plasma in $ \tau $--approach.
However, in work \cite {Kliewer} it is shown, that  expression
received in \cite {Lind} is noncorrect,
as does not turn into classical expression under a condition, when
quantum amendments can be neglected. Thus in work \cite {Kliewer}
empirically corrected expres\-sion  for
dielectric permeability of quantum plasma, free from
the specified lack has been offered.
By means of this expression the authors investigated
quantum amendments to optical properties of metal
\cite {Kliewer2}, \cite {Kliewer3}.

Dielectric permeability of quantum plasma  is widely
used also for studying the screening of the electric field and
Friedel oscillations (see, for example, \cite {Kohn1} - \cite {Harr}).
In the work \cite {Eminov}  screening of the Coulomb fields
in magnetised electronic gas has been is studied.

In the theory of quantum plasma there exist two essentially
different possibilities of construction of the relaxation kinetic
equation in $ \tau $ -- approximation: in the space of momentum (in the
space of Fourier images of the distribution function) and in the
space of coordinates. On the basis of the relaxation kinetic
equations in the space of momentum Mermin \cite {Mermin}
has carried out consistent derivation of the dielectric
permeability for quantum collisional plasma in 1970 for the first time.

In the present work expression for the longitudinal
dielectric permeability
with use of the relaxation equations in space of coordinates is deduced.
If in the obtained expression we make
Planck constant converges  to zero ($ \hbar\to 0$), we will receive
exactly classical expression of dielectric permeability of
non--degenerate plasma. Various limiting cases of the
dielectric permeability are investigated.
Comparison with Mermin's result is carried out also.

\section{solution of the kinetic equation}

We consider the kinetic Wigner --- Vlasov --- Boltzmann
equation \cite{Gurov} with collisional integral in the form
BGK--model
$$
\frac{\partial f}{\partial t}+\textbf{v}\frac{\partial f}{\partial
\textbf{r}}=\dfrac{ie}{\hbar}W[U]+\nu[f_{eq}-f].
\eqno{(2.1)}
$$

This equation describes evolution of the Wigner function
for electrons in quantum plasma.

Here $e$ is the charge of electron, $\nu$ is the effective collision
frequency of electrons with ions and neutral atoms,
$f(\mathbf{r}, \mathbf{p},t)$ is the Wigner function for electrons.
The function
$$
f_{eq}(\mathbf{r},v,t)=
\dfrac{1}{1+\exp\Big[\dfrac{mv^2}{2k_BT}-\dfrac{\mu(\mathbf{r},t)}
{k_BT}\Big]},
$$
is the equilibrium distribution Fermi --- Dirac function for
electrons,  $W[f]$ is the Wigner --- Vlasov functional
  for the scalar potential $U=U(\mathbf{r},t)$
$$
W[f]=\dfrac{1}{(2\pi)^3}\int \Big[U(\mathbf{r}-\dfrac{\hbar
\mathbf{b}}{2},t)\Big]-U(\mathbf{r}+\dfrac{\hbar
\mathbf{b}}{2},t)\Big]\times $$$$ \times f(\mathbf{r},\mathbf{p'},t)
e^{i\mathbf{b}(\mathbf{p'}-\mathbf{p})}\,d^3bd^3{p\,}',
\eqno{(2.2)}
$$
$\mathbf{b}=\{b_x,b_y,b_z\}$ is the vector,
$\hbar$ is the Planck constant, $k_B$ is the Boltzmann constant,
$\mu(\mathbf{r},t)$ is the chemical potential, $\mathbf{p}$ is
the momentum of electrons, $m$ and $\mathbf{v}$ are their mass and
velocity.

The Wigner function is analogue of distribution function  for quantum
systems. It is widely used in the various  physics
problems. Wigner's function was investigated, for example,
in works \cite {Arnold} and \cite {Kozlov}.

We introduce the Fourier transformation of the Wigner function
$$
F(\mathbf{r},\mathbf{b},t)=\dfrac{1}{(2\pi)^3}\int f(\mathbf{r},
\mathbf{p'},t)e^{i \mathbf{b}\mathbf{p'}}\,d^3{p\,}'.
$$

The Wigner --- Vlasov functional with the Fourier transformation
equals to
$$
W[f]=\int \Big[U\Big(\mathbf{r}-\frac{\hbar \mathbf{b}}{2},t\Big)-
U\Big(\mathbf{r}+\frac{\hbar \mathbf{b}}{2},t\Big)\Big]
F(\mathbf{r}, \mathbf{b},t)e^{-i\mathbf{b}\mathbf{p}}d^3p.
$$

Let's consider, that electron distribution  function depends on one
spatial coordinate $x$, time $t$ and momentum $\mathbf {p}$,
and the electric scalar potential depends on one spatial coordinate
$x$ and time $t$. We take the scalar potential in the form
$$
U(x,t)=U_0 e^{i(kx-\omega t)}.
\eqno{(2.3)}
$$

We will calculate the Wigner --- Vlasov functional (2.2).
It is easy to see, that
$$
U\Big(x-\frac{\hbar b_x}{2},t\Big)-
U\Big(x+\frac{\hbar b_x}{2},t\Big)=$$$$=U(x,t)\Big[\exp\Big(-
i\dfrac{\hbar k b_x}{2}\Big)-\exp\Big(i\dfrac{\hbar k b_x}{2}\Big)
\Big]
$$

Let's calculate internal integral in (2.2).
We receive, that
$$
\dfrac{1}{(2\pi)^3}\int \Big[U\Big(x-\frac{\hbar b_x}{2},t\Big)-
U\Big(x+\frac{\hbar b_x}{2},t\Big)\Big]\exp(i\mathbf{b}(
\mathbf{p'}-\mathbf{p}))\,d^3b=
$$
$$
=\dfrac{U(x,t)}{(2\pi)^3}\int \Big[\exp\Big(-
i\dfrac{\hbar k b_x}{2}\Big)-
\exp\Big(i\dfrac{\hbar k b_x}{2}\Big)\Big]
\exp(i\mathbf{b}(
\mathbf{p'}-\mathbf{p}))\,d^3b=
$$
$$
=\delta(p'_y-p_y)\delta(p_z'-p_z)\Big[\delta\Big(p_x'-p_x-
\frac{\hbar k}{2}\Big)-\delta\Big(p_x'-p_x+
\frac{\hbar k}{2}\Big)\Big].
$$

By means of last equality  the Wigner --- Vlasov functional (2.2)
it is possible to present in the form
$$
W[f]=U(x,t)\int \delta\big(p_{y}-p_{y}'\big)
\delta\big(p_{z}-p_{z}'\big)\times $$$$ \times
\Big[\delta\Big(p_x-p_x'+ \frac{\hbar k}{2}\Big)-
\delta\Big(p_x-p_x'- \frac{\hbar k}{2}\Big)\Big]f(x,\mathbf{p},t)\,d^3p'.
$$

Now it becomes clear, that
$$
W[f]=U(x,t)\Big[f^+(x,\mathbf{p},t)-f^-(x,\mathbf{p},t)\Big],
\eqno{(2.4)}
$$
where
$$
f^{\pm}=f(x,p_x\pm \dfrac{\hbar k}{2},p_y,p_z).
$$

The Fermi --- Dirac locally equilibrium  distribution $f_{eq}$
we will be linearize about absolute distribution of Fermi --- Dirac
$$
f_F(v)\equiv f_0(v)=\dfrac{1}{1+\exp\Big[\dfrac{mv_T^2}{2k_BT}-
\dfrac{\mu}{k_BT}\Big]}, \qquad \mu=\const.
$$

We will enter dimensionless electron velocity  and chemical potential
$$
\mathbf{c}=\dfrac{v}{v_0}, \qquad
v_0=\dfrac{1}{\sqrt{\beta}}=\sqrt{\dfrac{2k_BT}{m}},\qquad
\alpha(x)=\dfrac{\mu(x)}{k_BT}.
$$

Expressions for $f_{eq}$ and $f_0$ thus become simpler
$$
f_{eq}=\dfrac{1}{1+e^{c^2-\alpha(x)}}, \qquad
f_0(c)=\dfrac{1}{1+e^{c^2-\alpha}}.
$$
Here
$$
\alpha(x)=\alpha+\delta \alpha(x), \qquad \alpha=\const.
$$

Linearization of $f_{eq}$ leads us to expression
$$
f_{eq}=f_0(c)+g(c)\delta \alpha(x), \qquad
g(c)=\dfrac{e^{c^2-\alpha}}{(1+e^{c^2-\alpha})^2}.
\eqno{(2.5)}
$$

Now we search  the Wigner function in the form
$$
f=f_0(c)+U(x,t)g(c)h(\mathbf{c}).
\eqno{(2.6)}
$$

From the law of  number of particles conservation
$$
\int (f_{eq}-f)d\Omega_F=0, \;\qquad
d\Omega=\frac{2d^3p}{(2\pi\hbar)^3},
$$
we seek that
$$
\delta \alpha(x)=\dfrac{U(x,t)}{2\pi\varphi_0(\alpha)}
\int h(\mathbf{c})g(c)\,d^3c,
$$
where
$$
\varphi_0(\alpha)=\int\limits_{0}^{\infty}f_0(c)dc=
2\int\limits_{0}^{\infty}g(c)c^2dc.
$$

We will replace the Wigner function $f$ in Wigner --- Vlasov functional
in linear approximation  on Fermi -- Dirac absolute distribution
$f_0(c)$. We will substitute in the equation (2.1) linear expressions (2.5),
(2.6) and $W[f]=W [f_0]=U[f_0^+-f_0^-]$.
As a result we receive the equation
$$
h(\mathbf{c})g(c)[1-i\omega\tau+ik_1c_x]=\dfrac{ie}{\hbar \nu}
[f_0^+(\mathbf{c})-f_0^-(\mathbf{c})]+\dfrac{g(c)}{2\pi \varphi_0(\alpha)}
\int h(\mathbf{c})g(c)d^3c.
$$

From this equation we find
$$
h(\mathbf{c})g(c)=
\Bigg[\dfrac{g(c)}{2\pi\varphi_0(\alpha)}\dfrac{A}{1-i\omega\tau+ik_1c_x}+
\dfrac{ie}{\hbar\nu}\dfrac{f_0^+(\mathbf{c})-f_0^-(\mathbf{c})}
{1-i\omega\tau+ik_1c_x}\Bigg],
\eqno{(2.7)}
$$
where
$$
A=\int h(\mathbf{c})g(c)d^3c,
\eqno{(2.8)}
$$
$$
f_0^{\pm}(\mathbf{c})=\dfrac{1}
{1+\exp\Big[(c_x\pm q/2)^2+c_y^2+c_z^2-\alpha\Big]},
\qquad
q=\frac{\hbar k}{mv_0}=\dfrac{k}{k_0}.
$$
$k_0\equiv {mv_0}/{\hbar}$ is the thermal wave number.

For finding  the constant $A$ we will substitute (2.7) in (2.8).
As result it is received, that
$$
A=\dfrac{ie}{\hbar \nu}\int \dfrac{f_0^+(\mathbf{c})-f_0^-(\mathbf{c})}
{1-i\omega\tau +ik_1c_x}d^3c+\dfrac{A}{2\pi \varphi_0(\alpha)}
\int \dfrac{g(c)\,d^3c}{1-i\omega\tau +ik_1c_x}.
$$

Let's calculate two internal integrals
$$
\int\limits_{-\infty}^{\infty}\int\limits_{-\infty}^{\infty}
\dfrac{e^{c^2-\alpha}dc_ydc_z}{(1+e^{c^2-\alpha})^2}=
2\pi \int\limits_{0}^{\infty}\dfrac{e^{c_x^2+r^2-\alpha}\,r\,dr}
{(1+e^{c_x^2+r^2-\alpha})^2}=\dfrac{\pi}{1+e^{c_x^2-\alpha}}=
\pi f_0(c_x),
$$
$$
\int\limits_{-\infty}^{\infty}\int\limits_{-\infty}^{\infty}
[f_0^+(\mathbf{c})-f_0^-(\mathbf{c})]dc_ydc_z=
$$
$$
=2\pi \int\limits_{0}^{\infty}\bigg[\dfrac{1}{1+e^{(c_x+q/2)^2+
r^2-\alpha}}-\dfrac{1}{1+e^{(c_x-q/2)^2+r^2-\alpha}}\bigg]\,r dr.
$$

We notice that
$$
\int\limits_{0}^{\infty}\dfrac{rdr}{1+e^{(c_x+q/2)^2+r^2-\alpha}}=
\int\limits_{0}^{\infty}\dfrac{e^{-(c_x+q/2)^2-r^2+\alpha}\,rdr}
{1+e^{-(c_x+q/2)^2-r^2+\alpha}}=
%$$
%$$
%=
\dfrac{1}{2}\ln[1+e^{\alpha-(c_x+q/2)^2}].
$$

We find now, that
$$
\int\limits_{-\infty}^{\infty}\int\limits_{-\infty}^{\infty}
[f_0^+(\mathbf{c})-f_0^-(\mathbf{c})]dc_ydc_z=
\pi[\ln(1+e^{\alpha-(c_x+q/2)^2})-\ln(1+e^{\alpha-(c_x-q/2)^2})]=
$$
$$
=\pi\ln\dfrac{1+e^{\alpha-(c_x+q/2)^2}}{1+e^{\alpha-(c_x-q/2)^2}}.
$$

Hence, the quantity $A$ is equal
$$
A=\dfrac{ie\pi}{\hbar
\nu}\dfrac{J^+-J^-}{1-T/2\varphi_0(\alpha)},
\eqno{(2.9)}
$$
where
$$
T\equiv T(\omega\tau,k_1,\alpha)=\int\limits_{-\infty}^{\infty}
\dfrac{f_0(t)\,dt}{1-i\omega\tau+ik_1t}=
\int\limits_{-\infty}^{\infty}
\dfrac{\,dt}{(1+e^{t^2-\alpha})(1-i\omega\tau+ik_1t)},
$$
$$
J^{\pm}\equiv J^{\pm}(\omega\tau,k_1,q,\alpha)=\int\limits_{-\infty}^{\infty}
\dfrac{\ln[1+e^{\alpha-(t\pm q/2)^2}]\,dt}{1-i\omega\tau+ik_1t}=
$$
$$
=\int\limits_{-\infty}^{\infty}\dfrac{\ln(1+e^{\alpha-t^2})}
{1-i\omega\tau+ik_1t\mp ik_1q/2}.
$$

\section{Conductivity and Permeability}

Let's consider a relationship between electric field and potential
$$
\mathbf{E}(x,t)=-{\rm grad}\; U(x,t),
$$
or
$$
\mathbf{E}(x,t)=-\Big\{\frac{\partial U(x,t)}{\partial x},0,0\Big\},
$$
and a continuity  equation for current and charge
densities
$$
\dfrac{\partial \rho}{\partial t}+
\dfrac{\partial j_x}{\partial x}=0.
$$

Here according to definition
of electric conductivity we may represent the current density in the form
$$
j_x= \sigma_l E_x=-\sigma_l \dfrac{\partial U}{\partial x}
=%$$$$=
-\sigma_l U_0ik e^{i(kx-\omega t)}=-\sigma_l ik U(x,t).
$$
Hence,
$$
\dfrac{\partial j_x}{\partial x}=\sigma_lk^2U(x,t).
$$

Taking into account  obvious equality for charge density
$$
\rho=e\int fd\Omega_F=
e\int [f_0(c)+U(x,t)g(c)h(\mathbf{c})]\,
\dfrac{2p_0^3 d^3c}{(2\pi \hbar)^3},
$$
we obtain
$$
\dfrac{\partial \rho}{\partial t}=-i\omega eU(x,t)\int
h(\mathbf{c})\dfrac{2p_0^3 d^3c}{(2\pi \hbar)^3}
=-iU(x,t)\dfrac{\omega e2p_0^3 }{(2\pi \hbar)^3}A.
$$

From the continuity equation  and the expressions for derivative of
 current and charge density, we find
$$
\sigma_lk^2U(x,t)=-\dfrac{\partial \rho}{\partial t}=
iU(x,t)\dfrac{\omega e2p_0^3 }{(2\pi \hbar)^3}A,
$$
whence we receive expression for longitudinal dielectric conductivity
$$
\sigma_l=i\dfrac{\omega e2p_0^3 }{(2\pi \hbar)^3k^2}A,
$$
or, with using (2.9), we receive
$$
\sigma_l=-\dfrac{2\pi e^2\omega p_0^3}{k^2(2\pi\hbar)^3\hbar \nu}
\cdot\dfrac{J^+-J^-}{1-T/2\varphi_0(\alpha)}.
\eqno{(3.1)}
$$

The number of particles is equal in the equilibrium condition to
$$
N^{(0)}=\int f_0(c)d\Omega_F=\dfrac{2p_0^3}{(2\pi \hbar)^3}
\int f_0(c)d^3c=\dfrac{p_0^3\varphi_2(\alpha)}{\pi^2\hbar^3},
\eqno{(3.2)}
$$
where
$$
\varphi_2(\alpha)=\int\limits_{0}^{\infty}c^2f_0(c)dc.
$$

Using the equality  (3.2), we may transform the
expression for longitudinal conductivity (3.1)
$$
\dfrac{\sigma_l}{\sigma_0}=-\dfrac{\omega}{\nu qk_14\varphi_2(\alpha)}
\cdot\dfrac{J^+-J^-}{1-T/2\varphi_0(\alpha)}.
\eqno{(3.3)}
$$

Let's overwrite the formula (3.3) in the obvious form
$$
\dfrac{\sigma_l}{\sigma_0}=\dfrac{\omega}{\nu qk_14\varphi_2(\alpha)}
\dfrac{\displaystyle\int\limits_{-\infty}^{\infty}
\ln
\dfrac{1+e^{\alpha-(t-q/2)^2}}{1+e^{\alpha-(t+q/2)^2}}
\dfrac{dt}{1-i\omega\tau+ik_1t}}
{1-\dfrac{1}{2\varphi_0(\alpha)}\displaystyle\int\limits_{-\infty}^{\infty}
\dfrac{f_0(t)dt}{1-i\omega\tau+ik_1t}}.
\eqno{(3.4)}
$$

%If in numerator of integral to present in the form of a difference,
%in each of the received integrals to make obvious replacement of a variable, then
%to calculate a difference of integrals it is as a result received
%the following formula
After some manipulations the last expression may be transformed to

$$
\dfrac{\sigma_l}{\sigma_0}=-i\dfrac{\omega}{\nu 4\varphi_2(\alpha)}
\dfrac{\displaystyle\int\limits_{-\infty}^{\infty}
\dfrac{\ln(1+e^{\alpha-t^2})dt}{(1-i\omega\tau+ik_1t)^2+
k_1^2q^2/4}}
{1-\dfrac{1}{2\varphi_0(\alpha)}\displaystyle\int\limits_{-\infty}^{\infty}
\dfrac{f_0(t)dt}{1-i\omega\tau+ik_1t}}.
\eqno{(3.5)}
$$

Let's enter the plasma (Langmuir) frequency
$$
\omega_p=\dfrac{4\pi e^2 N^{(0)}}{m}.
$$

Using the formula for standard conductivity $\sigma_0=$
$e^2N^{(0)}/m\nu$=$\omega_p^2/4\pi\nu$, we present the dielectric
permeability in the form: $\varepsilon_l=4\pi i\sigma_l/\omega$.
If we take conductivity according to (3.4) we will receive
$$
\varepsilon_l=1+\dfrac{i\omega_p^2}{\nu^2qk_14\varphi_2(\alpha)}
\dfrac{\displaystyle\int\limits_{-\infty}^{\infty}\ln
\dfrac{1+e^{\alpha-(t-q/2)^2}}{1+e^{\alpha-(t+q/2)^2}}
\dfrac{dt}{1-i\omega\tau+ik_1t}}
{1-\dfrac{1}{2\varphi_0(\alpha)}\displaystyle\int\limits_{-\infty}^{\infty}
\dfrac{f_0(t)dt}{1-i\omega\tau+ik_1t}}.
\eqno{(3.6)}
$$

If to take conductivity under the formula (3.5), for permeability
we receive the following expression
$$
\varepsilon_l=1+\dfrac{\omega_p^2}{\nu^24\varphi_2(\alpha)}
\dfrac{\displaystyle\int\limits_{-\infty}^{\infty}
\dfrac{\ln(1+e^{\alpha-t^2})dt}{(1-i\omega\tau+ik_1t)^2+
k_1^2q^2/4}}
{1-\dfrac{1}{2\varphi_0(\alpha)}\displaystyle\int\limits_{-\infty}^{\infty}
\dfrac{f_0(t)dt}{1-i\omega\tau+ik_1t}}.
\eqno{(3.7)}
$$

Let's enter dimensionless parametres
$$
z=x+iy=\dfrac{\omega+i \nu}{k_0v_0}, \quad
x=\dfrac{\omega}{k_0v_0},
\quad y=\dfrac{\nu}{k_0v_0}=\dfrac{1}{k_0l}=\dfrac{1}{k_{01}}.
$$
Here $k_{01}=k_0l$ is the themal dimensionless wave number.
By means of these parametres we have
$$
1-i\omega\tau=\dfrac{\nu-i\omega}{\nu}=-i\dfrac{z}{y}, \quad
k_{01}=\dfrac{1}{y},\quad k_1=\dfrac{q}{y}.
$$

Besides, we receive
$$
1-i\omega\tau+ik_1t=\dfrac{i}{y}(qt-z), \quad
(1-i\omega\tau+ik_1t)^2+k_1^2q^2/4=-\dfrac{1}{y^2}[(qt-z)^2-q^4/4].
$$

Let's overwrite last four formulas in dimensionless parametres.
We receive for electric conductivity accordingly two formulas
$$
\dfrac{\sigma_l}{\sigma_0}=-i\dfrac{xy}{q^24\varphi_2(\alpha)}
\dfrac{\displaystyle\int\limits_{-\infty}^{\infty}
\ln\dfrac{1+e^{\alpha-(t-q/2)^2}}{1+e^{\alpha-(t+q/2)^2}}
\dfrac{dt}{qt-z}}
{1+\dfrac{iy}{2\varphi_0(\alpha)}\displaystyle\int\limits_{-\infty}^{\infty}
\dfrac{f_0(t)dt}{qt-z}}
\eqno{(3.8)}
$$
è
$$
\dfrac{\sigma_l}{\sigma_0}=i\dfrac{xy}{4\varphi_2(\alpha)}
\dfrac{\displaystyle\int\limits_{-\infty}^{\infty}
\dfrac{\ln(1+e^{\alpha-t^2})dt}{(qt-z)^2-q^4/4}}
{1+\dfrac{iy}{2\varphi_0(\alpha)}\displaystyle\int\limits_{-\infty}^{\infty}
\dfrac{f_0(t)dt}{qt-z}}.
\eqno{(3.9)}
$$

Similarly for dielectric permeability we have
$$
\varepsilon_l=1+\dfrac{x_p^2}{q^24\varphi_2(\alpha)}
\dfrac{\displaystyle\int\limits_{-\infty}^{\infty}
\ln\dfrac{1+e^{\alpha-(t-q/2)^2}}{1+e^{\alpha-(t+q/2)^2}}
\dfrac{dt}{qt-z}}
{1+\dfrac{iy}{2\varphi_0(\alpha)}\displaystyle\int\limits_{-\infty}^{\infty}
\dfrac{f_0(t)dt}{qt-z}}
\eqno{(3.10)}
$$
è
$$
\varepsilon_l=1-\dfrac{x_p^2}{4\varphi_2(\alpha)}
\dfrac{\displaystyle\int\limits_{-\infty}^{\infty}
\dfrac{\ln(1+e^{\alpha-t^2})dt}{(qt-z)^2-q^4/4}}
{1+\dfrac{iy}{2\varphi_0(\alpha)}\displaystyle\int\limits_{-\infty}^{\infty}
\dfrac{f_0(t)dt}{qt-z}}.
\eqno{(3.11)}
$$

In formulas (3.10) and (3.11) $x_p $ is the dimensionless plasma
frequency, $x_p =\omega_p/k_0v_0$.

\section{Special cases of conductivity and permeability}

Let's consider the limit of conductivity and permeability at $\nu\to 0$,
i.e. when collisional plasma passes in non--collisional.
For this purpose we take the formula (3.4) and
let's transform it in appropriate manner
%having allocated frequency of collisions
$$
\sigma_l=\dfrac{e^2N^{(0)}\omega }{4\varphi_2(\alpha)\hbar k^2}
\dfrac{\int\limits_{-\infty}^{\infty}\ln\dfrac{1+e^{\alpha-(t-q/2)^2}}
{1+e^{\alpha-(t+q/2)^2}}\dfrac{dt}{\nu-i\omega+ikv_0t}}
{1-\dfrac{\nu}{2\varphi_2(\alpha)}\int\limits_{-\infty}^{\infty}
\dfrac{f_0(t)dt}{\nu-i\omega+ikv_0t}}
\eqno{(4.1)}
$$

Passing in the formula (4.1) to a limit at $\nu\to 0$, we receive
the
expression for conductivity in non--collisional plasma:
$$
\sigma_l^\circ=i\dfrac{e^2N^{(0)}\omega}{4\varphi_2(\alpha)\hbar k^2}
\int\limits_{-\infty}^{\infty}\ln\dfrac{1+e^{\alpha-(t-q/2)^2}}
{1+e^{\alpha-(t+q/2)^2}}\dfrac{dt}{\omega-kv_0t}.
\eqno{(4.2)}
$$

Arguing in the same way, from the formula (3.5) in the limit at $\nu\to 0$
we receive the expression for conductivity in
non--collisional plasma
$$
\sigma_l^\circ=i\dfrac{e^2N^{(0)}\omega}{m4\varphi_2(\alpha)}
\int\limits_{-\infty}^{\infty}\dfrac{\ln(1+e^{\alpha-t^2})dt}
{(\omega-kv_0t)^2-k^2v_0^2q^2/4}.
\eqno{(4.3)}
$$

The formula (4.3) may be obtained  by the another way.
For this purpose the logarithm
in (4.2) we will write down in the form of  difference, and
we receive the difference of two integrals.
In each of these integrals we will make suitable replacement
variable. Then after simple transformations we come to
(4.3).

On the basis of (4.2) and (4.3) we will receive two expressions for
dielectric permeability in non--collisional plasma
$$
\varepsilon_l^\circ=1+\dfrac{\omega_p^2m}{\hbar k^24\varphi_2(\alpha)}
\int\limits_{-\infty}^{\infty}\ln\dfrac{1+e^{\alpha-(t+q/2)^2}}
{1+e^{\alpha-(t-q/2)^2}}\dfrac{dt}{\omega-kv_0t}
\eqno{(4.4)}
$$
and
$$
\varepsilon_l^\circ=1-\dfrac{\omega_p^2}{4\varphi_2(\alpha)}
\int\limits_{-\infty}^{\infty}\dfrac{\ln(1+e^{\alpha-t^2})dt}
{(\omega-kv_0t)^2-(k^2\hbar/2m)^2}.
\eqno{(4.5)}
$$

Formulas for calculation of the dielectric permeability in
non--collisional plasma is called in the literature by
Lindhard's dielectric functions.

Let's transform the formula (3.6) (or (3.7)) to a form convenient for
researches. We will give some representations of the dielectric
functions, convenient for research in various problems.
Let's enter dimensionless parameter
$w=\dfrac{\omega+i \nu}{kv_0}=\dfrac{z}{q}$. It
 is obvious, that $1-i\omega\tau+ik_1t=ik_1(t-w)$.
Hence,
$$
T=\int\limits_{-\infty}^{\infty}\dfrac{f_0(t)dt}{1-i\omega\tau+ik_1t}=
\dfrac{1}{ik_1}\int\limits_{-\infty}^{\infty}\dfrac{f_0(t)dt}{t-w}=
-\dfrac{i}{kl}\int\limits_{-\infty}^{\infty}\dfrac{f_0(t)dt}{t-w}=
$$
$$
=-\dfrac{i \nu}{kv_0}F_0(w),
$$
where
$$
F_0(w)=\int\limits_{-\infty}^{\infty}\dfrac{f_0(t)dt}{t-w}.
$$
Therefore we obtain the denominator from the formula (3.10) in
the following form
$$
1-\dfrac{1}{2\varphi_0(\alpha)}\int\limits_{-\infty}^{\infty}
\dfrac{f_0(t)dt}{1-i\omega\tau+ik_1t}=1+\dfrac{i
\nu}{kv_02\varphi_0(\alpha)}F_0(w).
$$

In the same way we will transform numerator from (3.6)
$$
\displaystyle\int\limits_{-\infty}^{\infty}
\dfrac{\ln(1+e^{\alpha-(t-q/2)^2})-
\ln(1+e^{\alpha-(t+q/2)^2})}{1-i\omega\tau+ik_1t}dt=
$$
$$
=\dfrac{1}{k_1^2}\int\limits_{-\infty}^{\infty}
\dfrac{\ln(1+e^{\alpha-t^2})dt}{(t-w)^2-q^2/4}.
$$

Therefore for dielectric permeability we obtain
$$
\varepsilon_l=1-\dfrac{\omega_p^2}{k^2v_0^24\varphi_2(\alpha)}
\dfrac{\int\limits_{-\infty}^{\infty}
\dfrac{\ln(1+e^{\alpha-t^2})dt}{(t-w)^2-q^2/4}}
{1+ivF_0(w)/2\varphi_0(\alpha)},
\eqno{(4.6)}
$$
where we have  put temporarily
$v=\dfrac{\nu}{kv_0}, w=u+iv, u=\dfrac{\omega}{kv_0}$.

Let's consider the denominator of the formula (4.6) and we will
transform it  in the following way
$$
1+\dfrac{ivF_0(w)}{2\varphi_0(\alpha)}=
1+\dfrac{ivwF_0(w)}{(u+iv)2\varphi_0(\alpha)}=1+\dfrac{i \nu wF_0(w)}
{(\omega+i \nu)2\varphi_0(\alpha)}=
$$
$$
=\dfrac{\omega+i \nu[1+wF_0(w)/2\varphi_0(\alpha)]}{\omega+i
\nu}.
$$

We will enter dispersion function which we name dispersion
Fermi --- Dirac function
$$
\lambda_0(w,\alpha)=1+\dfrac{w}{2\varphi_0(\alpha)}F_0(w)=
1+\dfrac{w}{2\varphi_0(\alpha)}
\int\limits_{-\infty}^{\infty}\dfrac{f_0(t)dt}{t-w}.
\eqno{(4.7)}
$$

Let's notice, that at $\alpha\to-\infty$
dispersion Fermi --- Dirac function
passes in dispersion function of classical plasma
$$
\lambda_c(w)=1+\dfrac{w}{\sqrt{\pi}}\int\limits_{-\infty}^{\infty}
\dfrac{e^{-t^2}dt}{t-w}.
$$

This function was introduced for the first time, apparently,
by Van Kampen \cite {Kampen}.

To establish this limiting transition, it is enough to notice,
that at $\alpha\to-\infty$ we have
$$
f_0(t)=\dfrac{1}{1+e^{\alpha-t^2}}\approx e^\alpha\cdot
e^{-t^2},
$$
$$
2\varphi_0(\alpha)\int\limits_{0}^{\infty}\dfrac{dt}{1+e^{\alpha-t^2}}
\approx 2e^\alpha
\int\limits_{0}^{\infty}e^{-t^2}dt=\sqrt{\pi}e^\alpha.
$$

Let's notice, that Fermi --- Dirac and Van Kampen dispersion functions
can be written down accordingly in the form
$$
\lambda_0(w,\alpha)=\dfrac{1}{2\varphi_0(\alpha)}
\int\limits_{-\infty}^{\infty}\dfrac{tf_0(t)dt}{t-w},
$$
and
$$
\lambda_c(w)=\dfrac{1}{\sqrt{\pi}}\int\limits_{-\infty}^{\infty}
\dfrac{te^{-t^2}dt}{t-w}.
$$

Thus, from the formula (4.6) it is possible to present the denominator in
the form
$$
1+\dfrac{ivF_0(w)}{2\varphi_0(\alpha)}=
\dfrac{\omega+i \nu \lambda_0(w,\alpha)}{\omega+i \nu}=
\dfrac{u+i v\lambda_0(w,\alpha)}{u+iv}.
$$

By means of this formula we will write down the formula for the longitudinal
Permeability of quantum collisional plasma in the form
$$
\varepsilon_l=1-\dfrac{u_p^2}{4\varphi_2(\alpha)}\cdot
\dfrac{(u+iv)L(w,q,\alpha)}{u+i v \lambda_0(w,\alpha)}.
\eqno{(4.7)}
$$

Here we was  entered the function
$$
L(w,q,\alpha)=\int\limits_{-\infty}^{\infty}\dfrac{\ln(1+e^{\alpha-t^2})dt}
{(t-w)^2-q^2/4},
\eqno{(4.8)}
$$
and dimensionless plasma frequency $u_p=\dfrac{\omega_p}{kv_0}$.

Let's consider the case when quantum plasma passes in
classical, i.e. $\hbar\to 0$ or $q\to 0$. In this case
the formula (4.7) passes in the following form
$$
\varepsilon_l^{\rm classic}=1-\dfrac{u_p^2}{4\varphi_2(\alpha)}\cdot
\dfrac{u+iv}{u+iv\lambda_0(w,\alpha)}\int\limits_{-\infty}^{\infty}
\dfrac{\ln(1+e^{\alpha-t^2})}{(t-w)^2}dt.
\eqno{(4.8)}
$$

Calculating integral from (4.8) in parts, we receive, that
$$
\int\limits_{-\infty}^{\infty}\dfrac{\ln(1+e^{\alpha-t^2})}{(t-w)^2}dt=
-2\int\limits_{-\infty}^{\infty}\dfrac{tf_0(t)dt}{t-w}=
-4\varphi_0(\alpha)\lambda_0(w,\alpha).
$$

Hence, the formula (4.8) will be transformed to the form
$$
\varepsilon_l^{\rm classic}=
1+u_p^2\cdot\dfrac{\varphi_0(\alpha)}{\varphi_2(\alpha)}\cdot
\dfrac{(u+iv)\lambda_0(w,\alpha)}{u+iv\lambda_0(w,\alpha)}.
\eqno{(4.9)}
$$

The formula (4.9) in accuracy coincides with the known formula for
longitudinal permeability of classical plasma of the arbitrary
degeneration degree  of electron gas. We will write down this
formula in dimensionless parametres $z=x+iy, q $:
$$
\varepsilon_l^{\rm classic}=
1+\dfrac{x_p^2\varphi_0(\alpha)}{q^2\varphi_2(\alpha)}\cdot
\dfrac{(x+iy)\lambda_0(z/q,\alpha)}{x+iy\lambda_0(z/q,\alpha)},
\eqno{(4.10)}
$$
where
$$
\lambda_0(z/q,\alpha)=\dfrac{q}{2\varphi_0(\alpha)}
\int\limits_{-\infty}^{\infty}\dfrac{tf_0(t)dt}{qt-z}, \qquad
z=\dfrac{\omega+i \nu}{k_0v_0},\qquad q=\dfrac{k}{k_0},
$$
or
$$
\varepsilon_l^{\rm classic}=1+\dfrac{\omega_p^2\varphi_0(\alpha)}
{k^2v_0^2\varphi_2(\alpha)}\cdot
\dfrac{(\omega+i \nu)\lambda_0(w,\alpha)}
{\omega+i \nu\lambda_0(w,\alpha)}, \qquad w=\dfrac{\omega+i
\nu}{kv_0}.
%\eqno{()}
$$

Let's give some more representations of dielectric permeability.
Let's enter auxiliary functions
$$
l(w\mp q/2)=\int\limits_{-\infty}^{\infty}\ln(1+e^{\alpha-(t\mp q/2)^2})
\dfrac{dt}{t-w}=\int\limits_{-\infty}^{\infty}\dfrac{\ln(1+e^{\alpha-t^2})}
{t-(w\mp q/2)}dt.
$$

We can present the Formula (3.6) in the form
$$
\varepsilon_l=1+\dfrac{\omega_p^2(\omega+i \nu)}
{qk^2v_0^24\varphi_2(\alpha)}\dfrac{l(w-q/2)-l(w+q/2)}
{\omega+i \nu \lambda_0(w,\alpha)},
\eqno{(4.11)}
$$
or
$$
\varepsilon_l=1+\dfrac{\omega_p^2w^2}{q4\varphi_2(\alpha)}
\dfrac{l(w-q/2)-l(w+q/2)}{(\omega+i \nu)(\omega+i \nu
\lambda_0(w,\alpha))}.
\eqno{(4.12)}
$$

We will notice, that
$$
l(w-q/2)-l(w+q/2)=-q \int\limits_{-\infty}^{\infty}
\dfrac{\ln(1+e^{\alpha-t^2})dt}{(t-w)^2-q^2/4}
\equiv-qL(w,q,\alpha).
$$

Hence, the formula (4.12) can be presented in the form
$$
\varepsilon_l=1-\dfrac{\omega_p^2(\omega+i \nu)L(w,q,\alpha)}
{k^2v_0^24\varphi_2(\alpha)(\omega+i \nu \lambda_0(w,\alpha))}.
\eqno{(4.13)}
$$

\section{Comparison with Mermin's result}

Mermin \cite {Mermin} considered the kinetic
relaxation equation in $\tau$ -- approach in momentum space
for finding the general expression of dielectric permeability.

Mermin (see Mermin N.D. \cite {Mermin})  has been received
the general expression of dielectric function
$$
\varepsilon^M(\omega,k)=
1+\dfrac{(\omega+i \nu)\Big[\varepsilon^\circ
(\omega+i \nu,k)-1\Big]}
{\omega +i\nu \dfrac{\varepsilon^\circ(\omega+i \nu,k)-1}
{\varepsilon^\circ(0,k)-1}}.
\eqno{(5.1)}
$$

In the formula (5.1) the designation is entered:
$\varepsilon^\circ(\omega,k)$ is the so-called
Lindhard's dielectric function, i.e.
the dielectric function received for non--collisional
plasma, expression $\varepsilon^\circ(\omega+i\nu,k)$
means, that argument of Lindhard dielectric function  $\omega$
 is replaced formally on $\omega+i\nu$. According to (4.11)
Lindhard  function looks like
$$
\varepsilon^\circ_l(\omega,k)=1+\dfrac{u_p^2k_0}{k4\varphi_2(\alpha)}
\Big[l(\frac{\omega}{kv_0}-\frac{k}{2k_0})-
l(\frac{\omega}{kv_0}+\frac{k}{2k_0})\Big],
$$
or
$$
\varepsilon^\circ_l(\omega,q)=1+\dfrac{u_p^2}{q4\varphi_2(\alpha)}
\Big[l(\frac{\omega}{kv_0}-\frac{q}{2})-
l(\frac{\omega}{kv_0}+\frac{q}{2})\Big],\qquad
u_p=\dfrac{\omega_p}{kv_0}.
$$

From last equality we deduce the following two formulas
$$
\varepsilon_l^\circ(\omega+i \nu,q)-1=\dfrac{u_p^2}{q4\varphi_2(\alpha)}
\cdot
\Big[l(\frac{\omega+i \nu}{kv_0}-\frac{q}{2})-
l(\frac{\omega+i \nu}{kv_0}+\frac{q}{2})\Big],
\eqno{(5.2)}
$$
$$
\varepsilon_l^\circ(0,q)-1=
\dfrac{u_p^2}{q4\varphi_2(\alpha)}
\cdot\Big[l(-q/2)-l(q/2)\Big].
\eqno{(5.3)}
$$

Let's make the relation of the left parts of equalities (5.2) and (5.3).
We have
$$
\dfrac{\varepsilon_l^\circ(\omega+i \nu,q)-1}
{\varepsilon_l^\circ(0,q)-1}=
\dfrac{l(\frac{\omega+i \nu}{kv_0}-\frac{q}{2})-
l(\frac{\omega+i \nu}{kv_0}+\frac{q}{2})}{l(-q/2)-l(q/2)}.
\eqno{(5.4)}
$$

By means of equalities (5.2) and (5.4) we can write down Mermin's formula (5.1)
in the form
$$
\varepsilon_l=1+\dfrac{u_p^2(\omega+ i \nu)}{q4\varphi_2(\alpha)}\cdot
\dfrac{l(\frac{\omega+i \nu}{kv_0}-q/2)-
l(\frac{\omega+i \nu}{kv_0}+q/2)}{\omega+i \nu d},
\eqno{(5.5)}
$$
where
$$
d=d(\omega+i \nu,q)=\dfrac{\varepsilon_l^\circ(\omega+i \nu,q)-1}
{\varepsilon_l^\circ(0,q)-1}=
\dfrac{l(\frac{\omega+i \nu}{kv_0}-q/2)-
l(\frac{\omega+i \nu}{kv_0}+q/2)}{l(-q/2)-l(q/2)}.
$$

The formula (5.5) gives representation of dielectric function
obtained with the use of kinetic equation in the form of
relaxation $\tau$ -- models in momentum space.

Let's notice, that at $\omega=0$ (a low-frequency limit) the Mermin's
formula gives representation of dielectric function, which not
depends on collisional frequency of electrons and looks like:
$$
\varepsilon_l^M=1+\dfrac{\omega_p^2k_0}{k^3v_0^24\varphi_2(\alpha)}
\Big[l(-\frac{k}{2k_0})-l(\frac{k}{2k_0})\Big].
$$

At the same time from our formula (4.11) we receive expression for
the dielectric function, depending on collisional frequency of
electrons
$$
\varepsilon_l=1+\dfrac{\omega_p^2}{qk^2v_0^24\varphi_2(\alpha)}
\dfrac{l(\frac{i \nu}{kv_0}-\frac{k}{2k_0})-l(\frac{i \nu}{kv_0}-
\frac{k}{2k_0})}{1+\frac{i \nu}{kv_02\varphi_0(\alpha)}
F_0(\frac{i \nu}{kv_0})}.
$$

For comparison we will present the formula (4.11) for
dielectric function:
$$
\varepsilon_l=1+\dfrac{u_p^2(\omega+ i \nu)}{q4\varphi_2(\alpha)}\cdot
\dfrac{l(\frac{\omega+i \nu}{kv_0}-\alpha)-
l(\frac{\omega+i \nu}{kv_0})}{\omega+i \nu \lambda_0(w,\alpha)}.
\eqno{(5.6)}
$$

The formula (5.6) gives representation of dielectric function
that obtained with the use  of  BGK--equation.

The difference between formulas (5.6) and (5.5) consists of the replacement
  the function $d$  by function
$\lambda_0$.

It is possible to show, that at $\hbar\to 0$ both formulas give  the same results.
It means, that at transition from
quantum plasma to classical dielectric function,
received on the basis of the kinetic equation as in momentum space,
and in coordinates space, passes in  the same dielectric function.

For this purpose it is necessary to prove, that
$$
\lim\limits_{\hbar\to 0}d=\lambda_0.
\eqno{(5.7)}
$$

Let's present expression for function $d$ in the following form
$$
d=\dfrac{l(w-q/2)-l(w+q/2)}{q}\cdot
\dfrac{q}{l(-q/2)-l(q/2)}.
$$

Earlier it has been shown, that
$$
\lim\limits_{q\to 0}\dfrac{l(w-q/2)-l(w+q/2)}{q}=-\lim\limits_{q\to 0}
\int\limits_{-\infty}^{\infty}\dfrac{\ln(1+e^{\alpha-t^2})dt}
{(t-w)^2-q^2/4}=
$$
$$
=-\int\limits_{-\infty}^{\infty}\dfrac{\ln(1+e^{\alpha-t^2})}
{(t-w)^2}dt=2\int\limits_{-\infty}^{\infty}\dfrac{tf_0(t)dt}
{t-w}=4\varphi_0(\alpha)\lambda_0(w,\alpha).
\eqno{(5.8)}
$$

From this equality at $w=0$ it is received
$$
\lim\limits_{q\to 0}\dfrac{l(-q/2)-l(q/2)}
{q}=4\varphi_0(\alpha)\lambda_0(0,q)=4\varphi_0(\alpha).
\eqno{(5.9)}
$$

From equalities (5.8) and (5.9) we find, that
$$
\lim\limits_{q\to 0}\dfrac{l(w-q/2)-l(w+q/2)}
{l(-q/2)-l(q/2)}\equiv
\lim\limits_{q\to0}d=\lambda_0(w,\alpha).
%\eqno{(4.10)}
$$
As it was required to show.

Thus, both relaxation BGK -- equations and in coordinates
space, and in momentum space at $q\to 0$
lead to the same dielectric function.

\section{Conclusion}

In the present work the correct formula for calculation of
longitudinal electric conductivity and dielectric permeability
in the quantum collisinal plasma under arbitrary degeneration degree
 of electron gas  is deduced.
For this purpose the Wigner --- Vlasov --- Boltzmann kinetic
equation with collisional integral in the form of BGK--model
(Bhatnagar, Gross and Krook) in coordinate space is used.
Comparison with Lindhard's formula has been realized.

\end{document}